\documentclass[aps,superscriptaddress,twocolumn]{revtex4}
\usepackage{amsmath,amssymb}
\usepackage{graphics,graphicx}
\usepackage{dcolumn,bm}

\newcommand{\sinp}{\affiliation{Theoretical Condensed Matter Physics Division
and Centre for Applied Mathematics and Computational Science,\\
Saha Institute of Nuclear Physics, 1/AF Bidhannagar, Kolkata 700 064, India.}}
\newcommand{\gdc}{\affiliation{Physics Department, Gurudas College, Narkeldanga,
 Kolkata 700 054, India.}}

\begin{document}

\title{Two Fractal Overlap Time Series: Earthquakes and Market Crashes}

\author{Bikas K. Chakrabarti}
\email{bikask.chakrabarti@saha.ac.in}
\sinp
\author{Arnab Chatterjee} 
\email{arnab.chatterjee@saha.ac.in}
\sinp
\author{Pratip Bhattacharyya}
\sinp \gdc
\begin{abstract}
We find prominent similarities in the features of the time series 
for the (model earthquakes or) 
overlap of two Cantor sets when one set moves with uniform relative
velocity over the other and time series of stock prices. 
An anticipation method for some of the crashes have been proposed here,
based on these observations.
\end{abstract}
\maketitle

\section{Introduction}
\label{intro}

Capturing dynamical patterns of stock prices are
major challenges both  for epistemologists as well as for financial analysts
\cite{CCB:book}. The statistical properties of their (time) variations
or fluctuations \cite{CCB:book} are now well studied and
characterized (with established fractal properties), but are not very
useful for studying and anticipating their dynamics
in the market. Noting that a single fractal gives essentially a time
averaged picture, a minimal two-fractal overlap time series model was
introduced \cite{CCB:Chakrabarti:1999,CCB:Pradhan:2003,CCB:Pradhan:2004}
to capture the time series of earthquake magnitudes. 
We find that the same model can be used to mimic and study the
essential features of the time series of stock prices.

\section{The two fractal-overlap model of earthquake}
\label{sec:overlapmodel}
\begin{figure}[t]
\centering
\resizebox*{4cm}{!}{\includegraphics{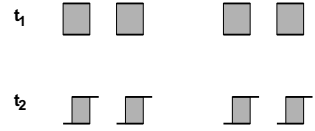}}
\caption{The overlap of two identical Cantor sets of
dimension $\ln 2/\ln 3$ at generation $n=2$ 
as one moves over the other with uniform
velocity. The total measure $O$ of the overlap (total shaded region)
varies with time and are shown for two different time instances.
}
\label{ccb:fig:generatio} 
\end{figure}

Let us consider first a geometric model~\cite{CCB:Chakrabarti:1999,CCB:Pradhan:2003,CCB:Pradhan:2004,CCB:Bhattacharyya:2005} 
of the
fault dynamics occurring in overlapping tectonic plates that form the earth's
lithosphere. A geological fault is created by a fracture in the earth's
rock layers followed by a displacement of one part relative to the other.
The two surfaces of the fault are known to be self-similar fractals. In
the model considered here~\cite{CCB:Chakrabarti:1999,CCB:Pradhan:2003,CCB:Pradhan:2004,CCB:Bhattacharyya:2005}, 
a fault is represented by a pair of overlapping identical
fractals and the fault dynamics arising out of the relative motion of the
associated tectonic plates is represented by sliding one of the fractals
over the other; the overlap $O$ between the two fractals represents the
energy released in an earthquake whereas $\log O$ represents the magnitude
of the earthquake. In the simplest form of the model each of the two
identical fractals is represented by a regular Cantor set of fractal
dimension $\log 2 / \log 3$ (see Fig.~\ref{ccb:fig:generatio}). 
This is the only exactly solvable model for
earthquakes known so far. The exact analysis of this model
\cite{CCB:Bhattacharyya:2005} for a finite generation $n$ of the Cantor sets
with periodic boundary conditions showed that the probability of
the overlap $O$, which assumes the values $O=2^{n-k} (k=0, \ldots , n)$,
follows the binomial distribution $F$ of $\log_2 O = n-k$ 
\cite{CCB:Bhattacharyya:2006}:
\begin{eqnarray}
\lefteqn{\Pr \left ( O=2^{n-k} \right )
 \equiv \Pr \left ( \log_2 O = n-k  \right )} \nonumber \\
 & &= \left ( \begin{array}{c} n\\ n-k \end{array} \right )
      \left ( {1 \over 3}  \right )^{n-k} \left ( {2 \over 3}  \right )^k
      \equiv F(n-k).
\label{eq:binomial-regular}
\end{eqnarray}
\noindent Since the index of the central term (i.e., the term for the most
probable event) of the above distribution is $n/3 + \delta$,
$-2/3 < \delta < 1/3$, for large values of $n$ Eq. (\ref{eq:binomial-regular})
may be written as
\begin{equation}
F \left ( {n \over 3} \pm r \right ) \approx \left ( \begin{array}{c}
                                            n\\ n \pm r \end{array} \right )
                            \left ( {1 \over 3}  \right )^{{n \over 3} \pm r}
                            \left ( {2 \over 3}  \right )^{{2n \over 3} \mp r}
\label{eq:binomial-central}
\end{equation}

\noindent by replacing $n-k$ with $n/3 \pm r$. For $r \ll n$,
we can write the normal approximation to the above binomial distribution as
\begin{equation}
F \left ( {n \over 3} \pm r  \right ) \sim {3 \over \sqrt{2 \pi n}}
                             \exp{ \left ( -{9r^2 \over 2n} \right )}
\label{eq:normal-approx}
\end{equation}

\noindent Since $\log_2 O = n-k = {n \over 3} \pm r$, we have
\begin{equation}
F \left ( \log_2 O \right ) \sim {1 \over \sqrt{n}}
          \exp{\left [ - {\left ( \log_2 O \right )^2 \over n} \right ]},
\label{eq:normal-approx'}
\end{equation}

\noindent not mentioning the factors that do not depend on $O$. Now
\begin{equation}
F \left ( \log_2 O \right ) \mathrm{d} \left ( \log_2 O \right )
\equiv G(O) \mathrm{d} O
\label{eq:equivalence}
\end{equation}
\noindent where
\begin{equation}
G(O) \sim {1 \over O}
 \exp \left [ - {\left ( \log_2 O \right )^2 \over n} \right ]
\label{eq:log-normal2}
\end{equation}

\noindent is the log-normal distribution of $O$. As the generation index
$n \to \infty$, the normal factor spreads indefinitely (since its width is
proportional to $\sqrt{n}$) and becomes a very weak function of $O$ so that
it may be considered to be almost constant; thus $G(O)$ asymptotically
assumes the form of a simple power law with an exponent that is
independent of the fractal dimension of the overlapping Cantor sets~\cite{CCB:Bhattacharyya:2006}:
\begin{equation}
G(O) \sim {1 \over O} \ \mathrm{for} \ n \to \infty.
\label{eq:power-law2}
\end{equation}

\section{The Cantor set overlap time series}
\label{sec:timeseries}
\begin{figure}[b]
\centering
\resizebox*{8.0cm}{!}{\includegraphics{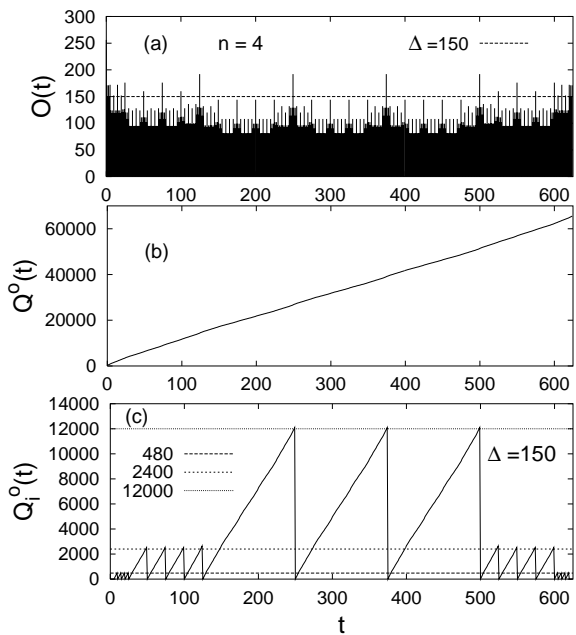}}
\caption{(a) The time series data of overlap size $O(t)$ for a regular Cantor
set of dimension $\rm{ln}4/\rm{ln}5$ at generation $n=4$.
(b) Cumulative overlap $Q^o(t)$ and
(c) the variation of the cumulative overlap $Q^o_i(t)$ for the same series,
where $Q$ is reset to zero after any big event of size
greater than $\Delta=150$. }
\label{ccb:figu1} 
\end{figure}

We now consider the time series $O(t)$ of the overlap set (of two identical
fractals~\cite{CCB:Pradhan:2004,CCB:Bhattacharyya:2005}), 
as one slides over the 
other with uniform velocity. Let us again consider two regular cantor sets
at finite generation $n$. As one set slides over the other, the overlap set
changes. The total overlap $O(t)$ at any instant
$t$ changes with time
(see Fig. \ref{ccb:figu1}(a)). In Fig. \ref{ccb:figu1}(b) we show the behavior of the cumulative overlap
\cite{CCB:Pradhan:2004} $Q^o(t) = \int_0^t O(\tilde{t}) d\tilde{t}$.
This curve, for sets with generation $n=4$,
is approximately a straight line \cite{CCB:Pradhan:2004} with slope $(16/5)^4$.
In general, this curve approaches a strict straight line in the limit
$a \rightarrow \infty$, asymptotically, where the overlap set comes from the
Cantor sets formed of $a-1$ blocks, taking away the central block,
giving dimension of the Cantor sets equal to $\mathrm{ln}(a-1)/\mathrm{ln}a$.
The cumulative curve is then almost a straight line and has then a slope
$\left[(a-1)^2/a\right]^n$ for sets of generation $n$.
If one defines a `crash' occurring at time $t_i$ when
$O(t_i)-O(t_{i+1}) \ge \Delta$ (a preassigned large value) and one
redefines the zero of the scale at each $t_i$,
then the behavior of the cumulative overlap
$Q^o_i(t) = \int_{t_{i-1}}^t O(\tilde t) d \tilde{t},\; \tilde{t} \le t_i$,
has got the peak value `quantization' as shown in 
Fig. \ref{ccb:figu1}(c). The reason is obvious. This justifies the simple thumb
rule: one can simply count the cumulative
$Q^o_i(t)$ of the overlaps since the last `crash' or `shock' at $t_{i-1}$
and if the value exceeds the minimum value ($q_o$), one can safely extrapolate
linearly and expect growth upto $\alpha q_o$ here and face a `crash' or overlap
greater than $\Delta$ ($=150$ in Fig.~\ref{ccb:figu1}). 
If nothing happens there,
one can again wait upto a time until which the cumulative grows upto 
$\alpha^{2}q_o$ and feel a `crash' and so on 
($\alpha=5$ in the set considered in Fig. \ref{ccb:figu1}).

\section{The stock price time series}
\label{sec:stocktimeseries}
\begin{figure}[b]
\centering
\resizebox*{8.0cm}{!}{\includegraphics{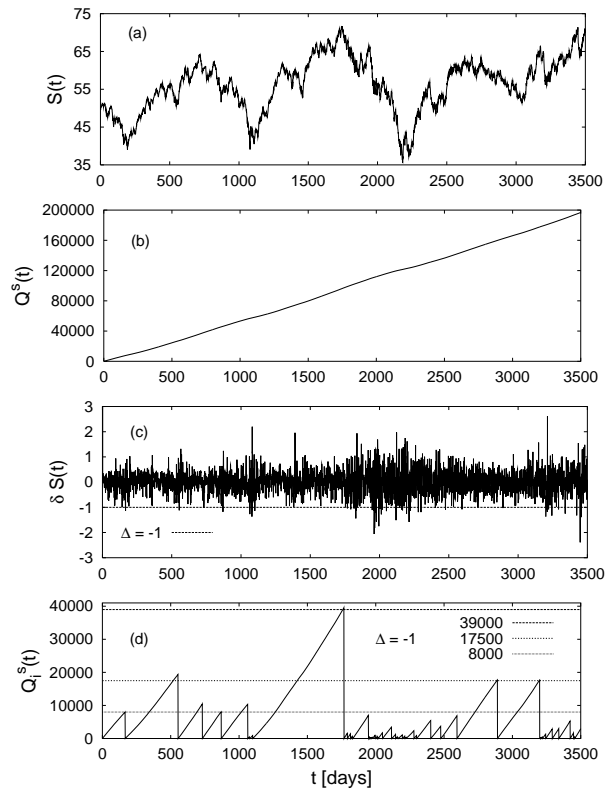}}
\caption{Data from New York Stock Exchange from January 1966 to December 1979:
industrial index \cite{CCB:NYSE}:
(a) Daily closing index $S(t)$ (b) integrated
$Q^s(t)$,
(c) daily changes $\delta S(t)$ of the index $S(t)$ defined as
$\delta S(t) = S(t+1) - S(t)$, and (d) behavior of $Q_i^s(t)$
where $\delta S(t_i) > \Delta$. Here, $\Delta=-1.0$ as shown in (c) by
the dotted line (from~\cite{CCB:PFE}).}
\label{ccb:figu2} 
\end{figure}
\begin{figure}
\centering
\resizebox*{8.0cm}{!}{\includegraphics{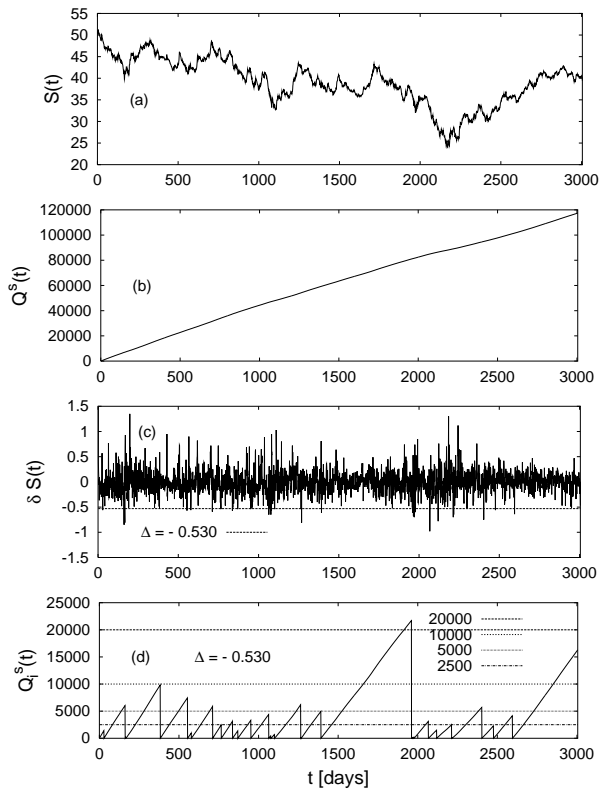}}
\caption{Data from New York Stock Exchange from January 1966 to December 1979:
utility index \cite{CCB:NYSE}:
(a) Daily closing index $S(t)$ (b) integrated
$Q^s(t)$,
(c) daily changes $\delta S(t)$ of the index $S(t)$ defined as
$\delta S(t) = S(t+1) - S(t)$, and (d) behavior of $Q_i^s(t)$
where $\delta S(t_i) > \Delta$. Here, $\Delta=-0.530$ as shown in (c) by
the dotted line.}
\label{ccb:figu3} 
\end{figure}

We now consider some typical stock price time-series data, available in the 
internet. The data analyzed here are for the New York Stock Exchange (NYSE)
Indices \cite{CCB:NYSE}.
In Fig. \ref{ccb:figu2}(a), we show that the daily stock price $S(t)$
variations for about $10$ years (daily closing price of the `industrial index')
from January 1966 to December 1979 (3505 trading days). The
cumulative $Q^s(t) = \int_0^t S(t) dt$ has again a straight line
variation with time $t$ (Fig. \ref{ccb:figu2}(b)).
Similar to the Cantor set analogy, we
then define the major shock by identifying those variations when
$\delta S(t)$ of the prices in successive days exceeded a preassigned
value $\Delta$ (Fig. \ref{ccb:figu2}(c)).
The variation of $Q_i^s(t) = \int_{t_{i-1}}^{t_i} S(\tilde{t}) d\tilde{t}$
where $t_i$ are the times when $\delta S(t_i) \le -1$ show similar
geometric series like peak values (see Fig. \ref{ccb:figu2}(d)); see
\cite{CCB:PFE}.

We observed striking similarity between the `crash' patterns in the
Cantor set overlap model and that derived from the data set of the
stock market index.
For both cases, the magnitude of crashes follow a similar pattern ---
the crahes occur in a geometric series.

A simple `anticipation strategy' for some of the crashes may be as follows:
If the cumulative $Q_i^s(t)$ since the last crash has grown beyond
$q_0 \simeq 8000$ here, wait until it grows (linearly with time) until about
$17,500$ ($\simeq 2.2q_0$) and expect a crash there. If nothing happens,
then wait until $Q_i^s(t)$ grows (again linearly with time) to a value of the 
order of $39,000$ ($\simeq (2.2)^2 q_0$) and expect a crash, and so on.

The same kind of analysis for the NYSE `utility index', for the same period,
is shown in Figs.~\ref{ccb:figu3}.

\section{Earthquake magnitude time series}
\begin{figure}[h]
\centering
\resizebox*{8.0cm}{!}{\includegraphics{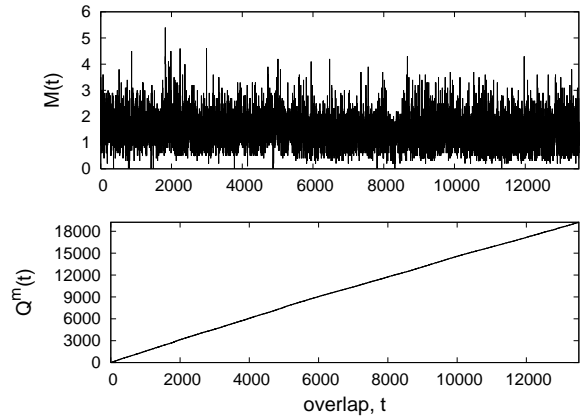}}
\caption{Local earthquake data from US geological Survey Southern
California catalogs.
\cite{CCB:USGS}:
\textit{Top}: Time series of successive quakes $M(t)$; 
\textit{Bottom}: integrated $Q^m(t)$.  
The dataset is a record of quakes between 1 January 2003 to 31 March 2004,
between depths 0 and 700 kms, between latitudes 32 N and  37 N and longitudes 
-122 W and -114 W.}
\label{ccb:figu4} 
\end{figure}

Unlike in the case of stock price time series where accurate data are
easily available, the time series for earthquake magnitudes $M(t)$ at any
fault involves considerably coordinated measurements and comparable
accuracies are not easily achievable. Still from the available data, 
as in the case of stock
market (where the integrated stock price $Q^s (t)$ shows clear linear 
variations with time and this fits well with that for the cumulative overlap 
$Q^o (t)$ for the fractal overlap model; see also \cite{CCB:EQbook}), 
the integrated earthquake magnitude 
$Q^m (t) = \int_0^t M(t) dt$ of the aftershocks does  also show
such prominent linear variations (see Fig.~\ref{ccb:figu4}). 
We believe, the slopes
of these linear $Q^m (t)$ vs. $t$ curves for different
faults would give us the signature of the corresponding fractal structure 
of the underlying fault. It may be noted in this context, in our model, the
slope becomes $[(a-1)^2/a]^n$ for an $n$th generation Cantor set,
formed out of the remaining $a-1$ blocks having the central block removed.

\section{Summary}
Based on the formal similarity between the two-fractal overlap model
of earthquake time series and of the stock market, we considered here a
detailed comparison. We find, the
features of the time series for the overlap of two Cantor sets when one 
set moves with uniform relative velocity over the other looks somewhat similar
to the time series of stock prices. 
We analyze both and explore the possibilities of anticipating a large
(change in Cantor set) overlap or a large change in stock price.
An anticipation method for some of the crashes has been proposed here,
based on these observations.




\begin{thebibliography}{99.}
\bibitem{CCB:book}
Sornette D (2003) Why Stock Markets Crash? Princeton Univ. Press, Princeton;
Mantegna RN, Stanley HE (1999) Introduction to Econophysics. 
Cambridge Univ. Press, Cambridge 

\bibitem{CCB:Chakrabarti:1999}
Chakrabarti BK, Stinchcombe RB (1999) Physica A 270:27-34

\bibitem{CCB:Pradhan:2003}
Pradhan S, Chakrabarti BK, Ray P, Dey MK (2003) Phys. Scr. T106:77-81

\bibitem{CCB:Pradhan:2004}
Pradhan S, Chaudhuri P, Chakrabarti BK (2004) in
Continuum Models and Discrete Systems, Ed. Bergman DJ, Inan E, 
Nato Sc. Series, Kluwer Academic Publishers, Dordrecht, pp.245-250;
cond-mat/0307735

\bibitem{CCB:Bhattacharyya:2005}
Bhattacharyya P (2005) Physica A 348:199-215 

\bibitem{CCB:Bhattacharyya:2006}
Bhattacharyya P, Chatterjee A, Chakrabarti BK (2007) 
Physica A, 381:377-382

\bibitem{CCB:NYSE}
NYSE Daily Index Closes from http://www.unifr.ch/econophysics


\bibitem{CCB:PFE}
Chakrabarti BK, Chatterjee A, Bhattacharyya P (2006)
in Takayasu H (Ed) Practical Fruits of Econophysics,
Springer, Tokyo, pp. 107-110; arxiv:physics/0510047.

\bibitem{CCB:EQbook}
Bhattacharyya P, Chakrabarti BK (Eds) (2006) 
Modelling Critical and Catastrophic Phenomena in Geoscience,
Lecture Notes in Physics, vol. 705, Springer-Verlag, Heidelberg

\bibitem{CCB:USGS}
U S Geological Survey, Southern California Catalogs, www.data.scec.org.

\end{thebibliography}
\end{document}